\documentclass[a4paper, 11pt]{article}
\usepackage{graphicx}
\usepackage{physics}
\usepackage{amsmath, amssymb, amsfonts, amsthm}
\usepackage{jheppub}
\usepackage{subcaption}
\newtheorem{theorem}{Theorem}

\begin{document}
\title{Tripartite entanglement in the HaPPY code is not holographic}
\author{Sriram Akella}
\emailAdd{sriram.akella@tifr.res.in}
\affiliation{Department of Theoretical Physics, Tata Institute of Fundamental Research, Homi Bhabha Road, Colaba, Mumbai 400005.}

\abstract{Holographic states satisfy several entropic inequalities owing to the Ryu-Takayangi formula. A drawback of these inequalities is that they only use bipartite entanglement in their formulation. We investigate a recently proposed ``GHZ-forbidding'' inequality, built out of the reflected entropy and the tripartite multi-entropy, that holds for holographic states. We show that the inequality is either violated or saturated, but never strictly satisfied, by stabilizer states, thereby showing that stabilizer states are not holographic. As a consequence, we show that tripartite entanglement in the HaPPY code is not holographic.}

\maketitle

\section{Introduction}\label{sec:Introduction}

\begin{figure}
\centering
 \includegraphics[width = 0.5\textwidth]{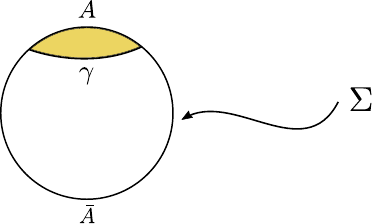}
 \caption{The RT surface $\gamma$ is the minimal area surface in the bulk that is homologous to $A$.}
 \label{fig:RT}
\end{figure}

The Ryu-Takayanagi formula \cite{Ryu:2006bv, Ryu:2006ef}, along with its covariant generalization \cite{Hubeny:2007xt}, is a hallmark of the AdS/CFT correspondence \cite{Maldacena:1997re, Witten:1998qj, Aharony:1999ti}. Suppose $\ket{\psi}$ is a boundary state in the CFT, prepared on a Cauchy slice $\Sigma$ partitioned into a subregion $A$ and its complement $\bar{A}$. The entanglement entropy of the state $\ket{\psi}$ with respect to this bipartition is
\begin{equation}
 S(A) = -\Tr \rho_A \log \rho_A = \frac{\mathcal{A}(\gamma)}{4G_N},
\end{equation}
where $\mathcal{A}(\gamma)$ is the area of the Ryu-Takayangi surface. The Ryu-Takayanagi surface is the minimal area surface homologous to $A$. See Fig. \ref{fig:RT} for an illustration. $\rho_A$ is the reduced density matrix of $\ket{\psi}$ on the Hilbert space $\mathcal{H}_A$ associated with the subregion $A$. 

The fact that entanglement entropy is given by the area of a minimal surface imposes stringent conditions on the entanglement structure of the holographic state $\ket{\psi}$. The Ryu-Takayanagi formula allows us to write down several entropic inequalities for the state $\ket{\psi}$. A well-known example of one such inequality is the strong subadditivity of entropy, which states that
\begin{equation}
 S(AB) + S(B C) \geq S(A  B C) + S(B).
\end{equation}
In the above inequality, we assume that the boundary is partitioned into at least three regions $A$, $B$, and $C$. For holographic states, this inequality follows from the minimality of the Ryu-Takayanagi surface as shown in \cite{Headrick:2007km}. Although strong subadditivity also holds for general quantum states, its proof is considerably more technical \cite{Lieb:1973cp}.

The Ryu-Takayanagi formula can be used to construct more complicated-looking inequalities which hold only for holographic states. For example, the tripartite mutual information is negative for holographic states \cite{Hayden:2011ag}: 
\begin{equation}
    S(A) + S(B) + S(C) - S(AB) - S(BC) - S(AC) + S(ABC) \leq 0,
\end{equation}
but not for general quantum states. The holographic entropy cone \cite{Bao:2015bfa, HernandezCuenca:2019wgh, Czech:2023xed} generates a plethora of such holographic inequalities. 

One drawback of these inequalities is that they use only bipartite entanglement entropy in their formulation. Therefore, these measures don't probe genuine features of multipartite entanglement. To illustrate this point with a simple example, consider the three-qubit W state and the following asymmetric GHZ state:
\begin{equation}
\begin{split}
 & \ket{W} = \frac{1}{\sqrt{3}}\ket{001} + \frac{1}{\sqrt{3}} \ket{010} + \frac{1}{\sqrt{3}}\ket{100}, \\ & \ket{aGHZ}  = \frac{1}{\sqrt{3}}\ket{000} + \sqrt{\frac{2}{3}}\ket{111}.
 \end{split}
\end{equation}
The reduced density matrices with respect to any bipartition are equal for these two states. Therefore, no quantity built out of bipartite measures of entanglement will distinguish the W from the asymmetric GHZ state. We might therefore suspect that the two states are equal up to a change of basis. However, there is no local unitary transformation that takes $\ket{W} \longleftrightarrow \ket{aGHZ}$; they belong to inequivalent LOCC classes \cite{Dur:2000zz}. We need genuine multipartite entanglement measures to distinguish such states. For all we know, holography might prefer the $W$ state over the asymmetric GHZ state, but inequalities coming from the holographic entropy cone can't tell them apart. 

In a recent paper \cite{Balasubramanian:2025hxg}, Balasubramanian and collaborators wrote down the first holographic entropic inequality using multipartite entanglement measures \cite{Horodecki:2024bgc, 2023arXiv230909459M, 2016arXiv161207747B, Szalay:2015vrx}. Their inequality takes the following simple form:
\begin{equation}\label{eq:ghz-forbid}
 \frac{1}{2}R^{(3)}(A: B) \geq GM^{(3)} (A: B: C) \quad \text{(for holographic states.)}
\end{equation}
The quantity on the left is called the \emph{residual information} \cite{Balasubramanian:2024ysu} or the \emph{Markov gap} \cite{Hayden:2021gno}. The quantity on the right is called the \emph{genuine tripartite multi-entropy} \cite{Iizuka:2025ioc, Iizuka:2025caq, Gadde:2022cqi}. We define both of these quantities in section \ref{sec2}, but the superscript means that these are tripartite measures of entanglement. The three-qubit (symmetric) GHZ state violates the above inequality, thereby showing that \emph{GHZ-like} entanglement is forbidden in holography. Because of this property, we will refer to Eq. (\ref{eq:ghz-forbid}) as the \emph{GHZ-forbidding inequality}.

In this paper, we will show that the above inequality is either violated or saturated for a class of quantum states called stabilizer states \cite{Gottesman:1997zz}. In terms of an equation, we show that
\begin{equation}
  \frac{1}{2}R^{(3)}(A:B) \leq GM^{(3)}(A: B: C) \quad \text{(for stabilizer states.)}
\end{equation}
We also show that the violation occurs if and only if the stabilizer state has genuine tripartite entanglement; the inequality is saturated otherwise. An interesting consequence of this result is that any toy model of holography that prepares a stabilizer state on the boundary fails to capture holographic features of tripartite entanglement.

The 3-qutrit code \cite{Cleve:1999qg} is a baby example often used as an analogy to explain how bulk degrees of freedom in AdS/CFT are encoded in the boundary \cite{Almheiri:2014lwa}. The code, however, happens to be a stabilizer code. It maps an unentangled state to a stabilizer state and, therefore, will violate the inequality. The more sophisticated HaPPY code \cite{Pastawski:2015qua} is also a stabilizer code which maps an unentangled bulk state\footnote{In fact, the argument goes through as long as the bulk state is itself a stabilizer state.} to a stabilizer state on the boundary, and cannot strictly satisfy the inequality.

It is crucial that the GHZ-forbidding inequality uses multipartite entanglement measures. We would not reach the above conclusion using bipartite entanglement measures alone. This is because the HaPPY code is designed to reproduce the minimal cut (Ryu-Takayanagi) formula for entanglement entropy. Therefore, it will satisfy all inequalities generated by the holographic entropy cone. The GHZ-forbidding inequality, however, is a stronger probe of the entanglement structure of holographic states. It allows us to rule out stabilizer states, like the ones prepared by the HaPPY code, as toy models of holographic states. 

The rest of the paper is organized as follows. In section \ref{sec2}, we define both sides of the GHZ-forbidding inequality and sketch the argument for why it holds for holographic states. In section \ref{sec3}, we evaluate the two sides of the GHZ-forbidding inequality for stabilizer states. The main result of this section is that the GHZ-forbidding inequality is either violated or saturated for stabilizer states. We apply our result to the HaPPY code \cite{Pastawski:2015qua} by formulating it as a stabilizer code \cite{Munne:2022ubc} in section \ref{sec4}. We conclude with a discussion in section \ref{sec5}.

\section{The GHZ-forbidding inequality}
\label{sec2}

In this section, we unpack both sides of the inequality in Eq. (\ref{eq:ghz-forbid}) and sketch a short proof coming from holography. The simpler quantity to explain is the one on the left. As mentioned in the introduction, it is called the residual information \cite{Balasubramanian:2024ysu} or the Markov gap \cite{Hayden:2021gno}. It is defined as the difference between the \emph{reflected entropy} \cite{Dutta:2019gen} and the mutual information:
\begin{equation}
 R^{(3)}(A:B) = S^{(R)}(A:B) - I(A:B).
\end{equation}
The mutual information, $I(A:B) = S(A) + S(B) - S(AB)$, is a quantity we all know and love. The definition of reflected entropy is more involved and goes as follows.

\subsection{Reflected entropy}
\begin{figure}
 \centering
 \includegraphics[width = 0.75\textwidth]{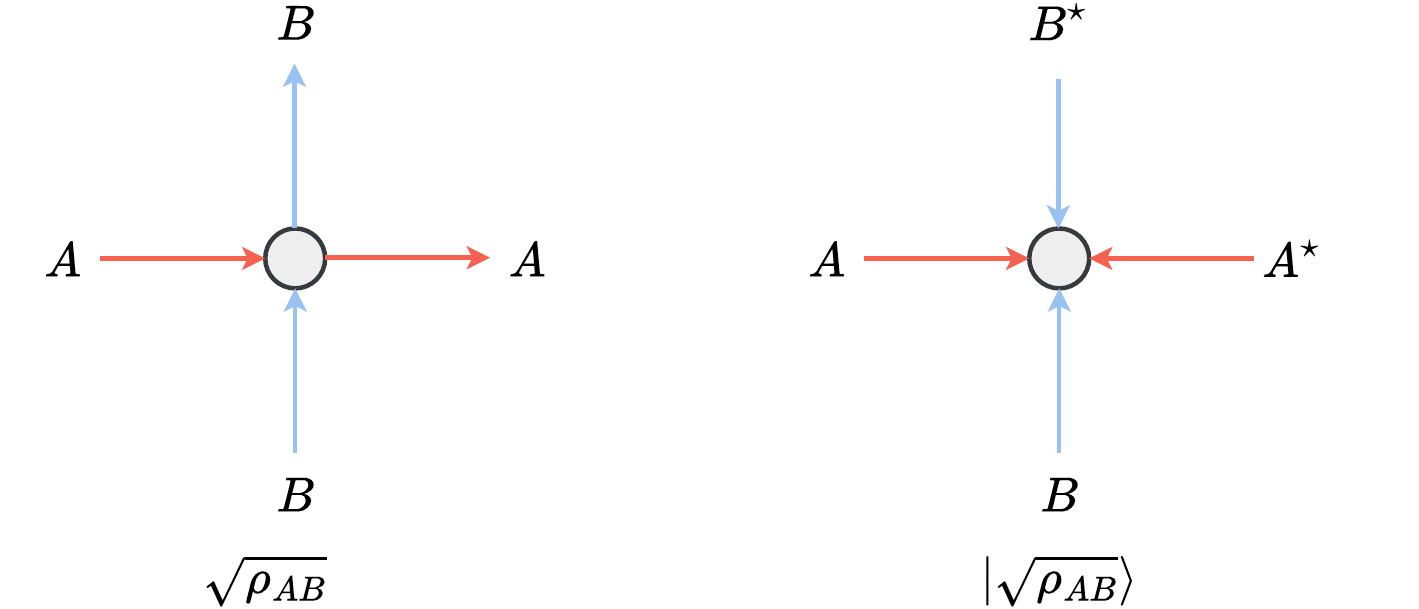}
 \caption{The operator $\sqrt{\rho_{AB}}$ takes $\mathcal{H}_A \otimes \mathcal{H}_B$ as input, and gives $\mathcal{H}_A \otimes \mathcal{H}_B$ as output. This is depicted with incoming $A$ and $B$ arrows and outgoing $A$ and $B$ arrows. The state $\ket{\sqrt{\rho_{AB}}}$, on the other hand, is depicted with four incoming arrows.}
 \label{fig:matrix-to-state}
\end{figure}

Consider a bipartite mixed state $\rho_{AB}$. Since $\rho_{AB}$ is a positive matrix, the square root, $\sqrt{\rho_{AB}}$, is a meaningful map from $\mathcal{H}_A \otimes \mathcal{H}_B \to \mathcal{H}_A \otimes \mathcal{H}_B$. We can view $\sqrt{\rho_{AB}}$ as a map from $\mathcal{H}_A \otimes \mathcal{H}_B \otimes \mathcal{H}_A^{\star} \otimes \mathcal{H}_B^{\star} \to \mathbb{C}$, where $\mathcal{H}_A^{\star}$ and $\mathcal{H}_B^{\star}$ are the dual Hilbert spaces of $\mathcal{H}_A$ and $\mathcal{H}_B$. Interpreted this way, $\sqrt{\rho_{AB}}$ becomes a state in the \emph{doubled} Hilbert space $\mathcal{H}_A \otimes \mathcal{H}_B \otimes \mathcal{H}_A^{\star} \otimes \mathcal{H}_B^{\star}$, and we denote it as $\ket{\sqrt{\rho_{AB}}}$. This is depicted in Fig. \ref{fig:matrix-to-state}. The pure state $\ket{\sqrt{\rho_{AB}}}$ is called the canonical purification of the mixed state $\rho_{AB}$. When $A^{\star}$ and $B^{\star}$ are traced over, we get $\rho_{AB}$ as the reduced density matrix on $A$ and $B$.

The most famous canonical purification perhaps is the thermofield double state \cite{, Takahashi:1996zn, Maldacena:2001kr, Maldacena:2013xja}. The mixed thermal state 
\[
\rho = \frac{1}{Z(\beta)}\sum_{E} e^{-\beta E} \ket{E}\bra{E}
\]
is canonically purified to the thermofield double state 
\[
\ket{TFD} = \frac{1}{\sqrt{Z(\beta)}} \sum_{E} e^{-\frac{\beta E}{2}} \ket{E} \otimes \ket{E}
\]
which lives in the doubled Hilbert space. The thermofield double state is unit normalized since $\Tr(\rho) = 1$, and its reduced density matrix on one of the constituent parties is the mixed thermal state.

As another example, consider a mixed two-qubit density matrix whose square root is
\begin{equation*}
 \sqrt{\rho_{AB}} = \frac{1}{\sqrt{2}}\ket{00} \bra{00} + \frac{1}{\sqrt{2}}\ket{11}\bra{11}.
\end{equation*}
To view this as a state in the doubled Hilbert space, we simply reverse the bras into kets (equivalent to flipping the arrows in Fig. \ref{fig:matrix-to-state}) to get
\begin{equation*}
 \ket{\sqrt{\rho_{AB}}} = \frac{1}{\sqrt{2}}\ket{0000} + \frac{1}{\sqrt{2}}\ket{1111}.
\end{equation*}
Since $\rho_{AB}$ has unit trace, this state is unit normalized.

Armed with the state $\ket{\sqrt{\rho_{AB}}}$, we can construct the following density matrix on $\mathcal{H}_A \otimes \mathcal{H}_A^{\star}$:
\begin{equation}
 \rho_{AA^{\star}} = \Tr_{BB^{\star}} \ket{\sqrt{\rho_{AB}}} \bra{\sqrt{\rho_{AB}}}.
\end{equation}
This mixed state is quite different from $\rho_{AB}$ which is obtained by tracing out $A^{\star}$ and $B^{\star}$. The reflected entropy is defined as the von Neumann entropy of the mixed state $\rho_{AA^{\star}}$, i.e,
\begin{equation}
 S^{(R)}(A:B) = - \Tr_{AA^{\star}} \left(\rho_{AA^{\star}} \log \rho_{AA^{\star}}\right).
\end{equation}
If $\rho_{AB}$ is pure, then the reflected entropy $S^{(R)}(A:B)$ is twice the entanglement entropy $S(A)$. 

The reflected entropy satisfies $I(A:B) \leq S^{(R)}(A:B) \leq 2 \min\{S(A), S(B)\}$. The Markov gap or residual information captures the difference between the reflected entropy and the mutual information: $R^{(3)}(A:B) = S^{(R)}(A:B) - I(A:B)$. Reflected entropy also admits a holographic dual as we soon discuss. Let us now turn our attention to the quantity that appears on the right side of the GHZ-forbidding inequality in Eq. (\ref{eq:ghz-forbid}).

\subsection{Genuine multi-entropy} 

As mentioned in the introduction, the quantity on the right is called the genuine tripartite multi-entropy \cite{Iizuka:2025ioc, Iizuka:2025caq}. It is defined as
\begin{equation}
 GM^{(3)}(A: B: C) = S^{(3)}(A: B: C) - \frac{1}{2}(S(A)+S(B) + S(C)),
\end{equation}
where $S^{(3)}(A: B: C)$ is the tripartite multi-entropy \cite{Gadde:2022cqi}. It is a multipartite entanglement measure defined via the replica trick. Let's spend some time discussing its definition and its key properties.

There is a graphical way of defining the multi-entropy as follows. Consider any $q$-partite state $\ket{\psi} \in \mathcal{H}_1 \otimes \dots \otimes \mathcal{H}_q$, along with a basis $\{\ket{\alpha_a}\}$ for each $\mathcal{H}_a$, and expand the state as
\begin{equation}
 \ket{\psi} = \sum_{\alpha_1 = 1}^{d_1} \dots \sum_{\alpha_q = 1}^{d_q} \psi_{\alpha_1 \dots \alpha_q} \ket{\alpha_1} \otimes \dots \otimes \ket{\alpha_q}.
\end{equation}
Each $\alpha_a$ runs from $1$ to $d_a$ which is the dimension of the Hilbert space $\mathcal{H}_a$. 

\begin{figure}
 \centering
 \begin{subfigure}{0.5\textwidth}
 \includegraphics[width = \textwidth]{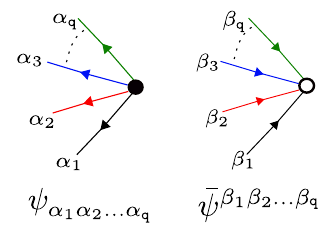}
  \caption{}
 \label{fig:psi}
 \end{subfigure}
 \hspace{0.075\textwidth}
\begin{subfigure}{0.4\textwidth}
     \includegraphics[width = \textwidth]{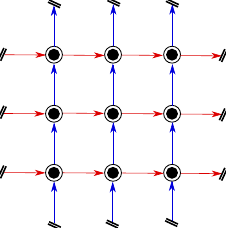}
        \caption{}
     \label{fig:square-grid}
\end{subfigure}
 \caption{(a) Graphical representation of a $q$-partite state $\ket{\psi} \in \mathcal{H}_1 \otimes \dots \otimes \mathcal{H}_q$ and its conjugate $\bra{\psi}$. (b) An example of the lattice with $n = 3$ sites in each of the $q = 2$ directions used to define the multi-entropy.}
\end{figure}

We can represent the coefficients $\psi_{\alpha_1\dots\alpha_q}$ as a tensor with $q$ outgoing colored legs, and the coefficients of the conjugate state $\bra{\psi}$ as a tensor with $q$ incoming colored legs as shown in Fig. \ref{fig:psi}. Each of the $q$ colors denotes one of the parties.

To define the multi-entropy, we imagine a $q$-dimensional cubical lattice where each site is occupied by a copy of $\psi$ and a copy of $\bar{\psi}$. There are $n$ sites in each of the $q$ directions giving a total of $n^q$ lattice sites. We impose periodic boundary conditions along each direction of the lattice, turning it into a $q$-dimensional torus. At each lattice site, we have $q$ incoming colored legs and $q$ outgoing colored legs corresponding to $\bar{\psi}$ and $\psi$.  These legs are imagined to be contracted along the edges of the lattice as described below.

Let us represent the sites of this lattice using vectors in $\mathbb{Z}_n^q$. The outgoing leg, corresponding to party $a$, at the lattice site $\vec{r} = (r_1, \dots, r_q)$, is contracted with the incoming party $a$ leg at the lattice site
\begin{equation}
 \sigma_a(\vec{r}) = (r_1, \dots, r_a + 1, \dots r_q).
\end{equation}
The addition above is modulo $n$. An example of these contractions for $q = 2$ and $n = 3$ is shown in Fig. \ref{fig:square-grid}. The slanted lines denote periodic boundary conditions.

Define the R\'enyi multi-entropy, $\mathcal{Z}_n^{(q)}$, to be the quantity obtained by this construction after contracting all the legs. The multi-entropy is defined as the following $n \to 1$ limit:
\begin{equation}\label{eq:renyi-multi}
 S^{(q)} = \lim_{n \to 1} \frac{1}{(1-n) n} \log \mathcal{Z}_n^{(q)}.
\end{equation}
It satisfies the following properties:
\begin{itemize}
 \item \textbf{Local unitary invariance.} The multi-entropy is a local unitary invariant. If $\ket{\psi}$ and $\ket{\phi}$ are two $q$-partite states related by a local unitary of the form $U_1 \otimes \dots U_q$, where each $U_a$ is a unitary operator on $\mathcal{H}_a$, then their multi-entropies are equal.

 \item \textbf{Symmetric in all parties.} The multi-entropy is symmetric in all the $q$ parties. In other words, it is invariant under any permutation of the $q$ parties.

 \item \textbf{Generalization of entanglement entropy.} When the number of parties, $q$, equals $2$, the multi-entropy reduces to the entanglement entropy.
\end{itemize}

There is an alternative  definition of the multi-entropy that requires a lesser number of index contractions. Consider a $(q-1)$-dimensional cubical lattice with $n$ sites in each direction and periodic boundary conditions. Instead of placing a copy of $\psi$ and $\bar{\psi}$ at each site, we place the reduced density matrix obtained by tracing one of the parties; say, the $q^{\text{th}}$ one. This density matrix has components
\begin{equation}
 \rho_{\alpha_1 \dots \alpha_{q-1}}^{\beta_1 \dots \beta_{q-1}} = \sum_{\alpha_q = 1}^{d_q} \psi_{\alpha_1 \dots \alpha_{q-1}\alpha_q}\bar{\psi}^{\beta_1 \dots \beta_{q-1} \alpha_q}.
\end{equation}

Interpreted graphically, this reduced density matrix has $(q-1)$ incoming colored legs and $(q-1)$ outgoing colored legs. These legs are again contracted along the edges of the lattice. If we call the quantity computed by this lattice as $\hat{\mathcal{Z}}_n^{(q-1)}$, then the multi-entropy is defined as
\begin{equation}
 S^{(q)} = \lim_{n \to 1} \frac{1}{1-n} \log \hat{\mathcal{Z}}_n^{(q-1)}.
\end{equation}
As shown in \cite{Gadde:2022cqi}, the above two definitions are equivalent. In terms of an equation,
\begin{equation}
\mathcal{Z}_n^{(q)}(\psi) = \left(\hat{\mathcal{Z}}_n^{(q-1)}(\rho)\right)^n
\end{equation}
for any $q$-partite state $\ket{\psi}$ and any of its $(q-1)$-partite reduced density matrices $\rho$. This follows from the symmetry properties of multi-entropy which are not important for our purposes.

The multi-entropy for a tripartite state, $S^{(3)}$, is the quantity that appears on the right of the GHZ-forbidding inequality. For a tripartite pure state, the inequality can be rewritten in terms of the reflected entropy, multi-entropy, and entanglement entropy as
\begin{equation}
\label{eq:ghz-forbid-pure}
 \frac{1}{2}S^{(R)}(A:B) + S(C) \geq S^{(3)}(A:B:C).
\end{equation}
This is the form of the inequality we consider since we only deal with tripartite pure states in this paper. Having defined the two sides of the GHZ-forbidding inequality, we now turn to their holographic duals.

\subsection{Holographic duals}
\begin{figure}
    \centering
    \begin{subfigure}{0.3\textwidth}
    \includegraphics[width=\textwidth]{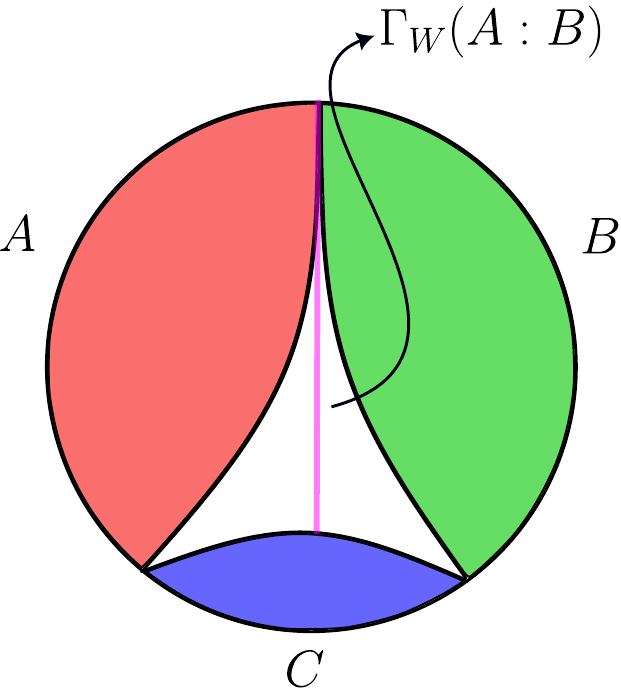}
    \caption{}
    \label{fig:three-ew}
    \end{subfigure}
    \hspace{0.2\textwidth}
    \begin{subfigure}{0.3\textwidth}
        \includegraphics[width=\textwidth]{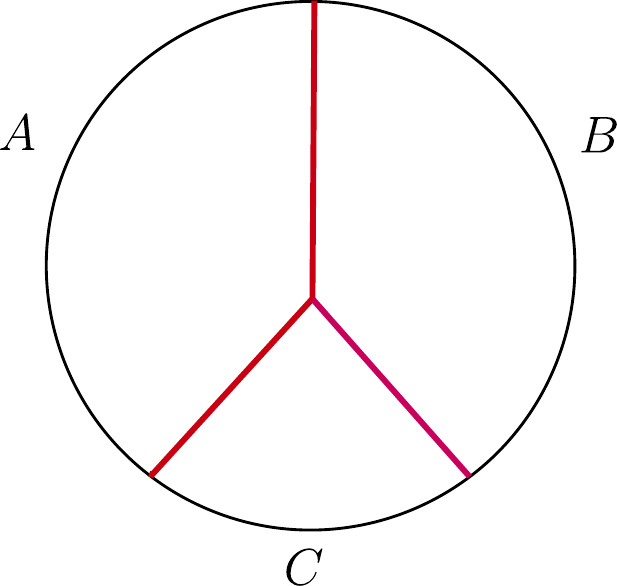}
    \caption{}
    \label{fig:3-multi-entropy}
    \end{subfigure}
    \caption{(a) The Ryu-Takayangi surfaces for a tripartite holographic state. The pink surface is the entanglement wedge cross-section surface. (b) The minimal area Mercedes-Benz surface whose area (divided by $4G_N$) is the tripartite multi-entropy for a holographic state.}
\end{figure}

For the purposes of illustration, let's consider vacuum AdS$_3$ so that the CFT lives on a cylinder. A Cauchy slice $\Sigma$ is a circle on this cylinder, and we partition it into three regions $A$, $B$, and $C$ as shown in Fig. \ref{fig:three-ew}. The areas (lengths in the case of AdS$_3$) of the Ryu-Takayanagi surfaces $\gamma(A)$, $\gamma(B)$, and $\gamma(C)$, divided by $4G_N$, are the entanglement entropies $S(A)$, $S(B)$, and $S(C)$.

For a holographic state, the reflected entropy $S^{(R)}(A:B)$ is twice the \emph{entanglement wedge cross-section} $E_W(A:B)$ \cite{Dutta:2019gen}. The entanglement wedge cross-section is the area of the minimal surface, $\Gamma_{W}(A:B)$, in the entanglement wedge of $A \cup B$ that separates $A$ and $B$. This is the pink curve in Fig. \ref{fig:three-ew}. 

The holographic dual of the tripartite multi-entropy $S^{(3)}(A:B:C)$ is given by minimizing the area of the Mercedes-Benz \cite{Gadde:2022cqi} shown in Fig. \ref{fig:3-multi-entropy}. To be more precise, decompose the bulk Cauchy slice into chambers such that:

\begin{itemize}
    \item each chamber is homologous to one of the boundary regions $A$, $B$, or $C$, and 
    \item the three chambers meet at a junction. 
\end{itemize}
Minimizing the area of this \emph{brane web} gives the tripartite multi-entropy $S^{(3)}(A: B: C)$ for a holographic state. 

These holographic duals are obtained via Euclidean gravity path integrals, following Lewkowycz-Maldacena \cite{Lewkowycz:2013nqa}. We assume that the Cauchy slice $\Sigma$ is time-reflection symmetric, and that the dominant gravitational saddle enjoys bulk replica symmetry. As argued in \cite{Penington:2022dhr}, there are situations where the bulk replica symmetry is broken when computing the tripartite multi-entropy $S^{(3)}(A:B:C)$. We will sweep this issue under the rug to derive the GHZ-forbidding inequality. However, there are multipartite entanglement measures that do preserve bulk replica symmetry \cite{Gadde:2024taa}. Holographic entropic inequalities built out of these measures won't suffer from this issue. 

Having explained the holographic duals of all the quantities that appear in the GHZ-forbidding inequality, we are now ready for the holographic proof. 

\subsection{Proof} 
Observe that the union of the entanglement wedge cross-section surface $\Gamma_W(A:B)$ and the Ryu-Takayanagi surface, $\gamma(C)$, of the boundary region $C$, gives a chamber decomposition of the bulk Cauchy slice satisfying the above listed properties. Therefore, we have
\begin{equation}
    E_W(A: B) + \text{Area}(\gamma(C)) \geq \mathcal{W}(A:B:C),
\end{equation}
where $\mathcal{W}(A:B:C)$ is the area of the minimal brane web (Mercedes-Benz). Simply dividing by $4G_N$, we obtain the GHZ-forbidding inequality as stated in Eq. (\ref{eq:ghz-forbid-pure}). 

The above argument also shows that the inequality is only saturated when the union of the Ryu-Takayanagi surface $\gamma(C)$ and the entanglement wedge cross-section is the minimal brane web. 

\section{Stabilizer formalism}
\label{sec3}

Now that we understand the GHZ-forbidding inequality and its proof for holographic states, let's turn to stabilizer states and discuss how to compute their reflected entropy and the tripartite multi-entropy. In this section, we prove the main result of this paper: stabilizer states can either violate or saturate the GHZ-forbidding inequality, but never strictly satisfy it. We begin with the definition of a stabilizer state \cite{Gottesman:1997zz}, followed by a discussion of their tripartite entanglement structure \cite{Bravyi_2006}. We write down explicit expressions for the reflected entropy and the tripartite multi-entropy for stabilizer states in Eq. (\ref{eq:ref-stab}). The main result is then an immediate consequence. 

\subsection{Stabilizer states}

Stabilizer states are a special class of many-qubit quantum states. They can also be defined for qudits, but we will stick to qubits for simplicity. They are special because of the Gottesman-Knill theorem \cite{Gottesman:1998hu} which states that stabilizer states can be simulated in polynomial time on a classical computer \cite{Aaronson:2004xuh}. Although simple, these states can be highly entangled \cite{Smith_2006} and demonstrate interesting multipartite entanglement structure \cite{Fattal:2004frh, Hein:2004zjp, Sharma:2024ypx}. As mentioned in the introduction, stabilizer tensor networks are often used as toy models of holography; the HaPPY code being a prime example.

An $n$-qubit stabilizer state $\ket{\psi}$ is defined as the unique state \emph{stabilized} by $n$ independent commuting elements $\{g_1, \dots, g_n\}$ of the $n$-qubit Pauli group $\mathcal{P}_n$. Recall that the Pauli group over $n$ qubits consists of all elements of the form
\begin{equation}
    \mathbf{P} = i^k P_1 \otimes \dots \otimes P_n, 
\end{equation}
where each $P_i$ takes values in $\{I, X, Y, Z\}$, and $k = 0, 1, 2, 3$. Here $X$, $Y$, $Z$ are the $2\times 2$ Pauli matrices thought of as operators acting on a qubit. If $\ket{\psi}$ is a stabilizer state, then 
\begin{equation}
    g_i \ket{\psi} = \ket{\psi}, \quad [g_i, g_j] = 0. 
\end{equation}
where each $g_i \in \mathcal{P}_n$. The $g_i$'s are called the generators of $\ket{\psi}$ and the group $G = \langle g_1, \dots, g_n\rangle$ generated by the $g_i$'s is called the stabilizer group of $\ket{\psi}$. The generators are independent in the sense that no generator can be written as a product of the other generators. It is easy to see that $G$ is isomorphic to $\mathbb{Z}_2^{n}$.

The Bell state 
\begin{equation}
    \ket{\phi^+} = \frac{1}{\sqrt{2}} \ket{00} + \frac{1}{\sqrt{2}}\ket{11}
\end{equation}
is a stabilizer states with generators $g_1 = XX$ and $g_2 = ZZ$. The stabilizer group of the Bell state is $G = \{II, XX, ZZ, -YY\}$. Another example of a stabilizer state is the GHZ state 
\begin{equation}
    \ket{GHZ} = \frac{1}{\sqrt{2}}\ket{000} + \frac{1}{\sqrt{2}}\ket{111}. 
\end{equation}
Its generators are $g_1 = XXX$, $g_2 = ZIZ$, $g_3 = IZZ$. An example of a state that is \emph{not} a stabilizer state is the W state: 
\begin{equation}\label{eq:W}
    \ket{W} = \frac{1}{\sqrt{3}}\ket{001} + \frac{1}{\sqrt{3}}\ket{010} + \frac{1}{\sqrt{3}}\ket{100}.
\end{equation}
The only element of the Pauli group $\mathcal{P}_3$ that stabilizes the W state is the identity $III$. 

Now, let us turn to understanding the tripartite entanglement structure of stabilizer states. Due to a remarkable theorem proved in \cite{Bravyi_2006}, the tripartite entanglement structure of stabilizer states can be completely characterized. We use this theorem to show that any tripartite entanglement measure, like the multi-entropy or the reflected entropy, can be explicitly written down for stabilizer states. The fact that stabilizer states violate the GHZ-forbidding inequality is then a direct consequence. 

\subsection{Tripartite entanglement}
Before we talk of tripartite entanglement, let us look at bipartite entanglement in stabilizer states. Given a stabilizer state $\ket{\psi}$, we can partition its qubits into $A$ and $B$, and compute the entanglement entropy. For a stabilizer state with a stabilizer group $G$, the entanglement entropy takes a  simple form \cite{Fattal:2004frh, Hein:2004zjp} given by\footnote{We will use $\log_2$ instead of the natural logarithm in the entropies since all of these subgroups are of the form $\mathbb{Z}_2^{p}$ for some integer $p$.}
\begin{equation}
    S(A) = S(B) = \frac{1}{2} \log_2 \left(\frac{|G|}{|G_A| |G_B|}\right), 
\end{equation}
where $|G|$ denotes the cardinality of the group $G$. The groups $G_A$ and $G_B$ are the subgroups of $G$ containing elements that act as identity on the qubits in $B$ and $A$ respectively. 

Now, partition the qubits of a stabilizer state $\ket{\psi}$ into three parties $A$, $B$, and $C$. We are interested in the reflected entropy $S^{(R)}(A:B)$ and the tripartite multi-entropy $S^{(3)}(A:B:C)$ for the state $\ket{\psi}$. The following GHZ extraction theorem \cite{Bravyi_2006} comes to our rescue:

\begin{theorem}[GHZ extraction]
    Any tripartite stabilizer state is local-unitary equivalent to a collection of: (a) GHZ states, (b) Bell pairs, and (c) unentangled qubits. 
\end{theorem}

We can even count the number of GHZ states, Bell pairs, and unentangled qubits that can be extracted from the state. For this we will need to define a few more subgroups of the stabilizer group $G$. As before, we have the subgroups $G_A$, $G_B$, $G_C$, defined as having trivial action outside $A$, $B$, and $C$ respectively. Similarly, we have the subgroups $G_{AB}$, $G_{BC}$, and $G_{AC}$, having trivial action on $C$, $A$, and $B$ respectively. We then define the following subgroup:
\begin{equation}
    G_{AB}\cdot G_{BC} = \{g \cdot h: g \in G_{AB}, h \in G_{BC}\}
\end{equation}
generated by taking products of elements in $G_{AB}$ and $G_{BC}$. Similarly, we have the subgroup $G_{AB}\cdot G_{BC}\cdot G_{AC}$ consisting of all elements of the form $f \cdot g \cdot h$, where $f \in G_{AB}$, $g \in G_{BC}$, and $h \in G_{AC}$.

The subgroup $G_{AB} \cdot G_{BC} \cdot G_{AC}$ is not necessarily the full stabilizer group $G$. As an example, consider the GHZ state again and think of $A$, $B$, and $C$ as containing the first, second, and third qubit respectively. The group $G_{AB}$ is generated by $ZZI$. Similarly, the groups $G_{BC}$ and $G_{AC}$ are generated by $IZZ$ and $ZIZ$ respectively. The group $G_{AB} \cdot G_{BC} \cdot G_{AC}$ is then generated by two elements: $IZZ$ and $ZIZ$. It automatically contains $ZZI$ since $IZZ \cdot ZIZ = ZZI$. However, the full stabilizer group is generated by three generators: $XXX$, $IZZ$, and $ZIZ$. Therefore, $G_{AB}\cdot G_{BC} \cdot G_{AC} \neq G$ for the GHZ state. 

The number of GHZ states that can be extracted from a general tripartite stabilizer state is given by this difference \cite{Bravyi_2006}:
\begin{equation}
    p = \log_2\left(\frac{|G|}{|G_{AB}\cdot G_{BC} \cdot G_{AC}|}\right).
\end{equation}
The number of Bell pairs extracted between $A$ and $B$ is 
\begin{equation}
    m_{AB} = \frac{1}{2} \left(\log_2\left(\frac{|G_{AB}|}{|G_A||G_B|}\right) - p\right),
\end{equation}
with similar expressions for $m_{BC}$ and $m_{AC}$. The number of unentangled states, by a simple counting, is\footnote{Getting to the second equality requires the following identity which we will state but not prove: $|G||G_A| |G_B| |G_C| = |G_{AB}| |G_{BC}| |G_{AC}|$.}
\begin{equation}
    s = n - 3p - 2(m_{AB} + m_{BC} + m_{AC}) = \log_2(|G_A| |G_B| |G_C|).
\end{equation}
What this theorem allows is to write any tripartite stabilizer state (up to local unitaries) as 
\begin{equation}
    \ket{\psi}_{ABC} \equiv \ket{GHZ}^{\otimes p} \otimes \ket{\phi^+}_{AB}^{\otimes m_{AB}} \otimes \ket{\phi^+}_{BC}^{\otimes m_{BC}} \otimes \ket{\phi^+}_{AC}^{\otimes m_{AC}} \otimes \ket{0}^{s}.
\end{equation}
We can now exploit this theorem to write down simple expressions for the reflected entropy and the multi-entropy. To do this we make the following rather obvious observations. 

The reflected entropy and the tripartite multi-entropy are local unitary invariant, and satisfy the following property: If $\ket{\psi}_{ABC} = \ket{\phi}_{A_1 B_1 C_1} \otimes \ket{\chi}_{A_2 B_2 C_2}$ is a tensor product of two states such that $A = A_1 \cup A_2$, etc., then both the reflected entropy and the multi-entropy satisfy
\begin{equation}
    \mathcal{E}(\ket{\psi}_{ABC}) = \mathcal{E}(\ket{\phi}_{A_1 B_1 C_1}) + \mathcal{E}(\ket{\chi}_{A_2 B_2 C_2}).
\end{equation}
In other words, the reflected entropy and tripartite multi-entropy are additive under the tensor products. Let's promote these two properties (along with a third property) to a definition.

A \textbf{\textit{tripartite entanglement invariant}} takes a tripartite pure state as an input and gives a positive real number as an output, such that:
\begin{itemize}
    \item \textbf{Local unitary invariance.} If $\ket{\psi}_{ABC}$ and $\ket{\phi}_{ABC}$ two states related via a local unitary of the form $U_A \otimes U_B \otimes U_C$, then $\mathcal{E}(\ket{\psi}) = \mathcal{E}(\ket{\phi})$. 

    \item \textbf{Additive under tensor products.} As explained above. 

    \item \textbf{Zero for product states.} If $\ket{\psi}_{ABC} = \ket{\phi}_A \otimes \ket{\chi}_B \otimes \ket{\xi}_C$ is a product state, then $\mathcal{E}(\ket{\psi}_{ABC}) = 0$. 
\end{itemize}
Both the reflected entropy and multi-entropy are tripartite entanglement invariants according to the above definition. 

Combining the definition of a tripartite entanglement invariant with the GHZ extraction theorem, we come to the following conclusion. Any tripartite entanglement invariant $\mathcal{E}$ for a stabilizer state can be calculated by calculating $\mathcal{E}$ for the GHZ state and the three different Bell states $\ket{\phi^+}_{AB}$, $\ket{\phi^+}_{BC}$, and $\ket{\phi^+}_{AC}$. 

A few simple calculations show that the reflected entropy and the tripartite multi-entropy for the GHZ \cite{Balasubramanian:2025hxg} and Bell states are as shown in Table \ref{tab:calc}.
Note that we are working with $\log_2$ instead of the natural logarithm, hence there are no factors of $\log(2)$.

\begingroup
\setlength{\tabcolsep}{12pt}
\setlength{\arrayrulewidth}{0.4mm}
\renewcommand{\arraystretch}{1.25}
\begin{table}
\centering 
\begin{tabular}{|c|c|c|}
\hline
      & $S^{(R)}(A:B)$ & $S^{(3)}(A:B:C)$   \\
     \hline 
     $\ket{GHZ}$ & 1 & 2 \\ 
     \hline 
     $\ket{\phi^{+}}_{AB}$ & 2 & 1 \\
     \hline 
     $\ket{\phi^{+}}_{AC}$ & 0 & 1 \\ 
     \hline
     $\ket{\phi^{+}}_{BC}$ & 0 & 1 \\
     \hline
\end{tabular}
\caption{The values of the reflected entropy $S^{(R)}(A:B)$ and the tripartite multi-entropy $S^{(3)}(A:B:C)$ for the GHZ state and the three Bell pairs.}
\label{tab:calc}
\end{table}
\endgroup 

Combining the results of these calculations with the theorem, the reflected entropy and the tripartite multi-entropy for a general tripartite stabilizer state are\footnote{In getting the expression for the multi-entropy we again have to appeal to the identity: $|G||G_A| |G_B| |G_C| = |G_{AB}| |G_{BC}| |G_{AC}|$.}
\begin{equation}\label{eq:ref-stab}
\begin{split}
    & S^{(R)}(A:B) = \log_2\left(\frac{|G_{AB}|}{|G_A| |G_B|}\right), \\
    & S^{(3)}(A:B:C) = \frac{1}{2}\log_2\left(\frac{|G|}{|G_A||G_B||G_C|}\right) + \frac{1}{2} \log_2\left(\frac{|G|}{|G_{AB} \cdot G_{BC} \cdot G_{AC}|}\right).
\end{split}
\end{equation}
From our discussion of the entanglement entropy for stabilizer states, we already have 
\begin{equation}
    S(C) = \frac{1}{2}\log_2 \left(\frac{|G|}{|G_C| |G_{AB}|}\right). 
\end{equation}

\subsection{Stabilizer states are not holographic}

We are now ready to analyze the GHZ-forbidding inequality in Eq. (\ref{eq:ghz-forbid-pure}) for stabilizer states. Plugging the above expressions in, we see some cancellations and are left with 
\begin{equation}
    |G_{AB} \cdot G_{BC} \cdot G_{AC}| \geq |G|\quad (!?) 
\end{equation}
This is a remarkable inequality because the left side, by definition, is a subgroup of the right side. Therefore, tripartite stabilizer states can never strictly satisfy the GHZ-forbidding inequality. Since the inequality was derived for holographic states, it follows that stabilizer states are not holographic.

This establishes the main result of the paper. Now we turn to applying this result to the HaPPY code. 

\section{Application to the HaPPY code}
\label{sec4}
In the previous section, we showed that stabilizer states are not holographic. Let's apply this result to the HaPPY code in this section. We consider two versions of the HaPPY code from \cite{Pastawski:2015qua}. We consider a tiling of the hyperbolic disk and place perfect tensors at the center of each tile. Perfect tensors are equivalent to absolutely maximally entangled (AME) states, and known AME states are stabilizer states (although there are exceptions we discuss in section \ref{sec5}). In the first version of the HaPPY code, all internal legs of the tensors are contracted to prepare a stabilizer state on the boundary. In the second version of the HaPPY code, one leg from each tile is left uncontracted to get a bulk-to-boundary stabilizer code. 

We begin by reviewing the theory of perfect tensors, which form the building blocks of the HaPPY code. To be explicit, we use the AME(6, 2) perfect tensor and the $\{6, 4\}$ tiling of the hyperbolic disk for the first version, and the $\{5, 4\}$ pentagon tiling for the second version, although the construction is more general. We show that the first version of the HaPPY code prepares a boundary stabilizer state and the second version is a stabilizer code from the bulk to the boundary. Therefore, a bulk stabilizer state is mapped to a boundary stabilizer state in the second version. In both cases, the boundary state is a stabilizer state and hence cannot reproduce holographic features of tripartite entanglement. Feel free to skip to the discussion in section \ref{sec5} if you are already convinced that the HaPPY code prepares stabilizer states on the boundary.

\subsection{Perfect Tensors} 

\begin{figure}
 \centering
 \includegraphics[width = 0.6\textwidth]{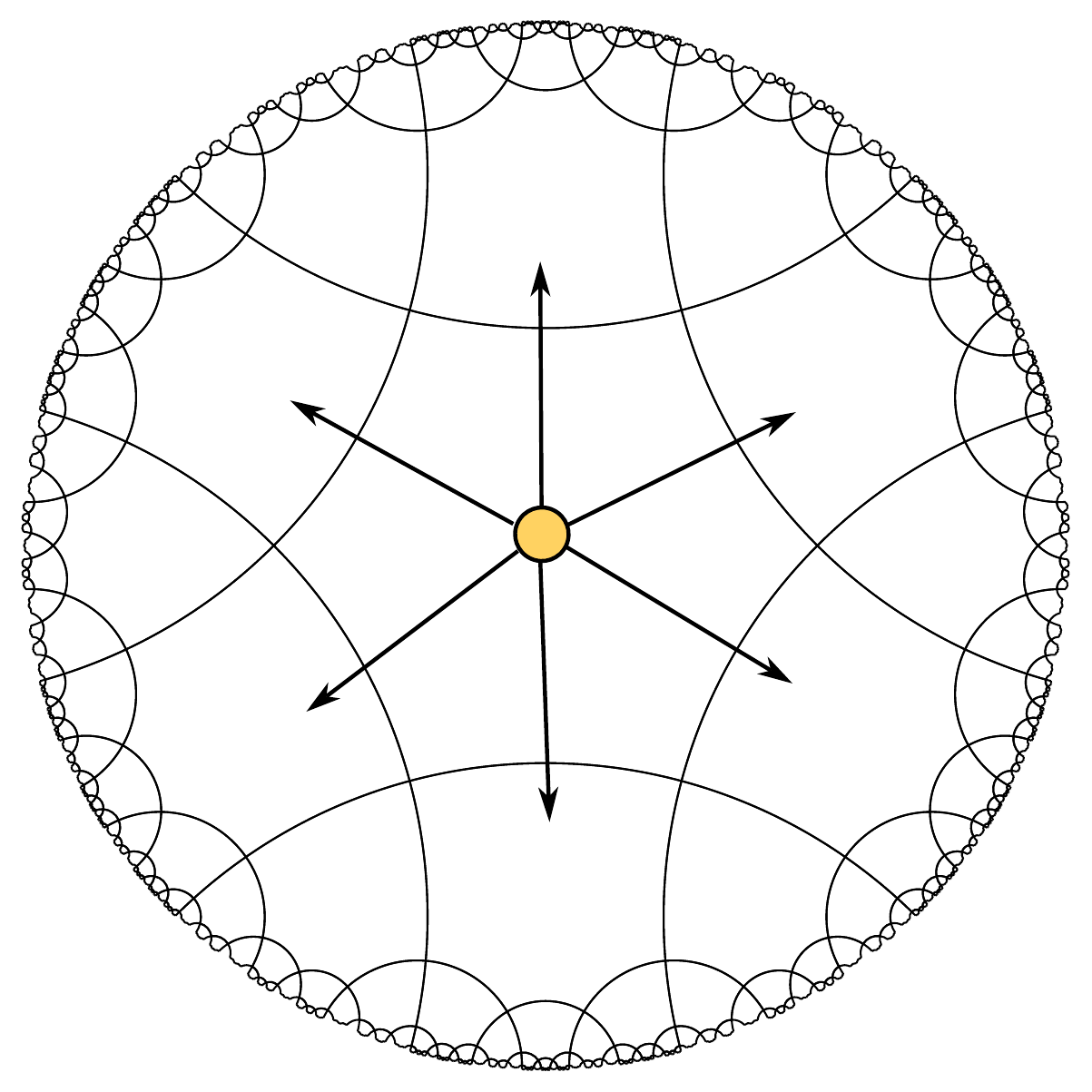}
 \caption{The $\{6, 4\}$ tiling of the hyperbolic disk with hexagons. We place a six-legged perfect tensor at the center of each hexagon.}
 \label{fig:hexagon}
\end{figure}

The HaPPY code is defined on some tiling of the hyperbolic disk. For example, we could take the $\{6, 4\}$ tiling and place a six-legged tensor at the center of each hexagon as shown in Fig. \ref{fig:hexagon}. The HaPPY code requires that this six-legged tensor be perfect as defined below.  

\begin{figure}
 \centering
 \includegraphics[width = 0.8\textwidth]{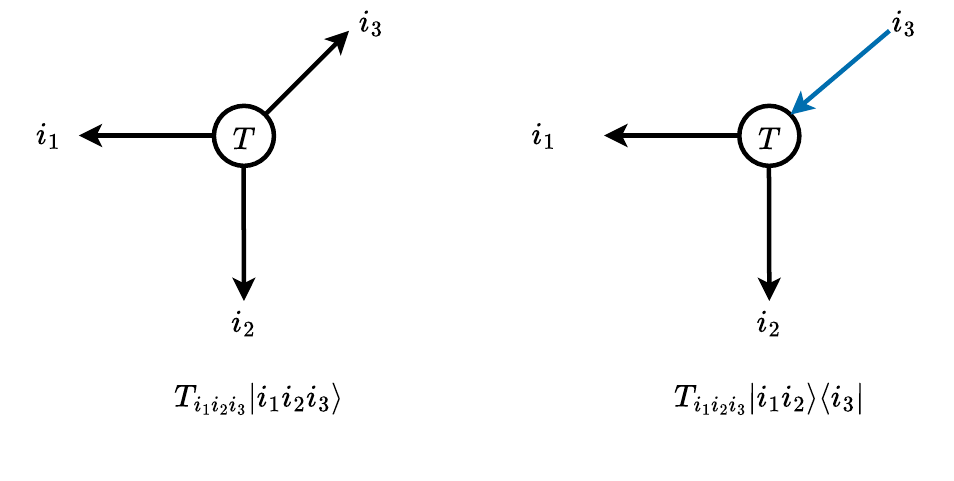}
 \caption{A tensor with legs going outwards can be thought of as a state. Reversing one of the legs, the tensor is now thought of as a map from one leg to two legs.}
 \label{fig:reverse-arrow}
\end{figure}

Given a tensor with $k$ legs, say, $T_{i_1 \dots i_k}$, consider the following unnormalized state 
\begin{equation}
    \ket{T} = \sum_{i_1 = 1}^{d_1} \dots \sum_{i_k=1}^{d_k} T_{i_1 \dots i_k} \ket{i_1 \dots i_k},
\end{equation}
where we associate a Hilbert space to each leg/index of the tensor $T$. If the leg carries an index $i_1$, then the associated Hilbert space is the span of $\{\ket{i_1}\}$ where $i_1$ takes values in the range $\{1, \dots, d_1\}$. When defining the reflected entropy, we turned bras into kets to go from a mixed state to a pure state. We can play the same game with the above unnormalized state, by turning kets into bras, to create a map from a subset of legs to the complementary subset of legs. As an example, if we turn $\ket{i_k}$ into $\bra{i_k}$, then we get the operator
\[
\sum_{i_1} \dots \sum_{i_k} T_{i_1 \dots i_k} \ket{i_1 \dots i_{k-1}} \bra{i_k}. 
\]
This is a map from the Hilbert space of the $k^{\text{th}}$ leg to a tensor product of the Hilbert spaces of the remaining $k-1$ legs. We represent this by flipping the orientation of the legs as shown in Fig. \ref{fig:reverse-arrow} for $k = 3$.

 Perfect tensors are even-legged tensors such that any partition of its legs into $A$ and $\bar{A}$, with $|A| \leq |\bar{A}|$, results in an isometry from $A \to \bar{A}$. Recall that $T: \mathcal{H}_A \to \mathcal{H}_{\bar{A}}$ is an isometry if 
\begin{equation}
    T^{\dagger}T = \mathbf{1}_A,
\end{equation}
where $\mathbf{1}_A$ is the identity operator on $\mathcal{H}_A$. It follows that the unnormalized state $\ket{T}$, corresponding to a perfect tensor, is \emph{absolutely maximally entangled (AME)} \cite{Helwig:2013ckb}, i.e., the reduced density matrix on any subset $A$ such that $|A| \leq |\bar{A}|$ is maximally mixed.

Most known AME states are stabilizer states \cite{Helwig:2013qoq}. Even in the HaPPY code \cite{Pastawski:2015qua}, the six-legged tensor that sits at each hexagon is taken to be the stabilizer state AME(6, 2) which is a six-qubit stabilizer state with generators
\begin{equation}
    \begin{split}
        g_1 & = X \otimes Z \otimes Z \otimes X \otimes I \otimes I, \\ 
        g_2 & = I \otimes X \otimes Z \otimes Z \otimes X \otimes I, \\ 
        g_3 & = X \otimes I \otimes X \otimes Z \otimes Z \otimes I, \\ 
        g_4 & = Z \otimes X \otimes I \otimes X \otimes Z \otimes I, \\
        g_5 & = X \otimes X \otimes X \otimes X \otimes X \otimes X,\\ 
        g_6 & = Z \otimes Z \otimes Z \otimes Z \otimes Z \otimes Z. 
    \end{split}
\end{equation}

We could consider more general tilings of the hyperbolic disk and place stabilizer AME states on each tile. Our results continue to apply in such situations. However, there are examples of AME states that are not stabilizer states \cite{Rather:2021vff, Rajchel-Mieldzioc:2025mqh, Casas:2025qpq}. Tensor networks prepared using non-stabilizer perfect tensors may reproduce features of holographic tripartite entanglement. To keep the discussion simple and explicit, we stick to the AME(6, 2) state.

\subsection{Building the network}

Having defined perfect tensors, we are now ready to prepare the state on the boundary. In the first version of the HaPPY code, we consider the $\{6, 4\}$ tiling of the hyperbolic disk and place the AME(6, 2) state at the center of each hexagon. When two hexagons share an edge, we contract the corresponding legs. For example, if $T_{i_1 i_2 i_3}$ is a three-legged tensor and $S_{j_1 j_2 j_3 j_4}$ is a four-legged tensor, then contracting the $i_1$ of $T$ with the $j_3$ of $S$ gives us the five-legged tensor
\begin{equation}
    (TS)_{i_2 i_3 j_1 j_2 j_4} = \sum_{i_1} \sum_{j_3} \delta^{i_1 j_3} T_{i_1 i_2 i_3} S_{j_1 j_2 j_3 j_4},
\end{equation}
which we depict as 
\begin{equation*}
    \includegraphics[width = 0.6\textwidth]{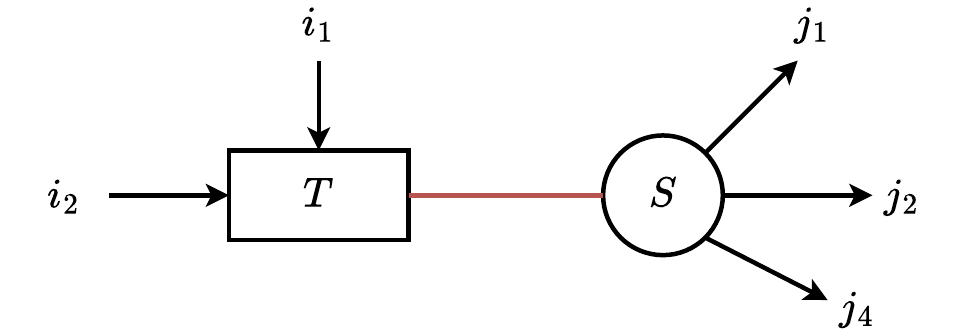}
\end{equation*}
For this expression to make sense, the range of $i_1$ and $j_3$ must be the same.

\begin{figure}
    \centering
    \includegraphics[width=0.5\linewidth]{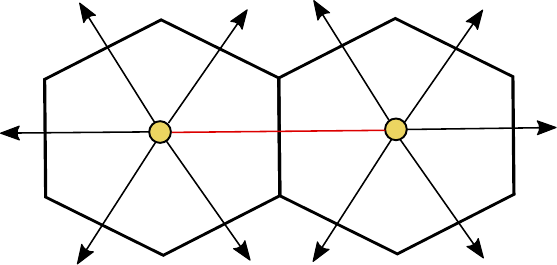}
    \caption{Contracting two legs cutting across a shared edge produces a state on the remaining ten legs.}
    \label{fig:hexagon-contract}
\end{figure}

In the case of the AME(6, 2) state, all legs are qubit legs taking values in $\{0, 1\}$. Any pair of legs can be contracted without a problem. Contracting two hexagons along a shared edge produces a ten-qubit state on the remaining uncontracted legs as shown in Fig. \ref{fig:hexagon-contract}.

Proceeding this way, we contract all the legs where two hexagons meet, and we produce a state on the boundary legs.  The boundary legs are those that cut across the boundary of the hyperbolic disk which is cutoff at some finite depth. This is an explicit realization of the first version HaPPY code. Although each tensor was perfect, the network obtained after contracting all legs is not. In the next subsection, we show that this boundary state is a stabilizer state. 

\begin{figure}
    \centering
    \includegraphics[width=0.5\linewidth]{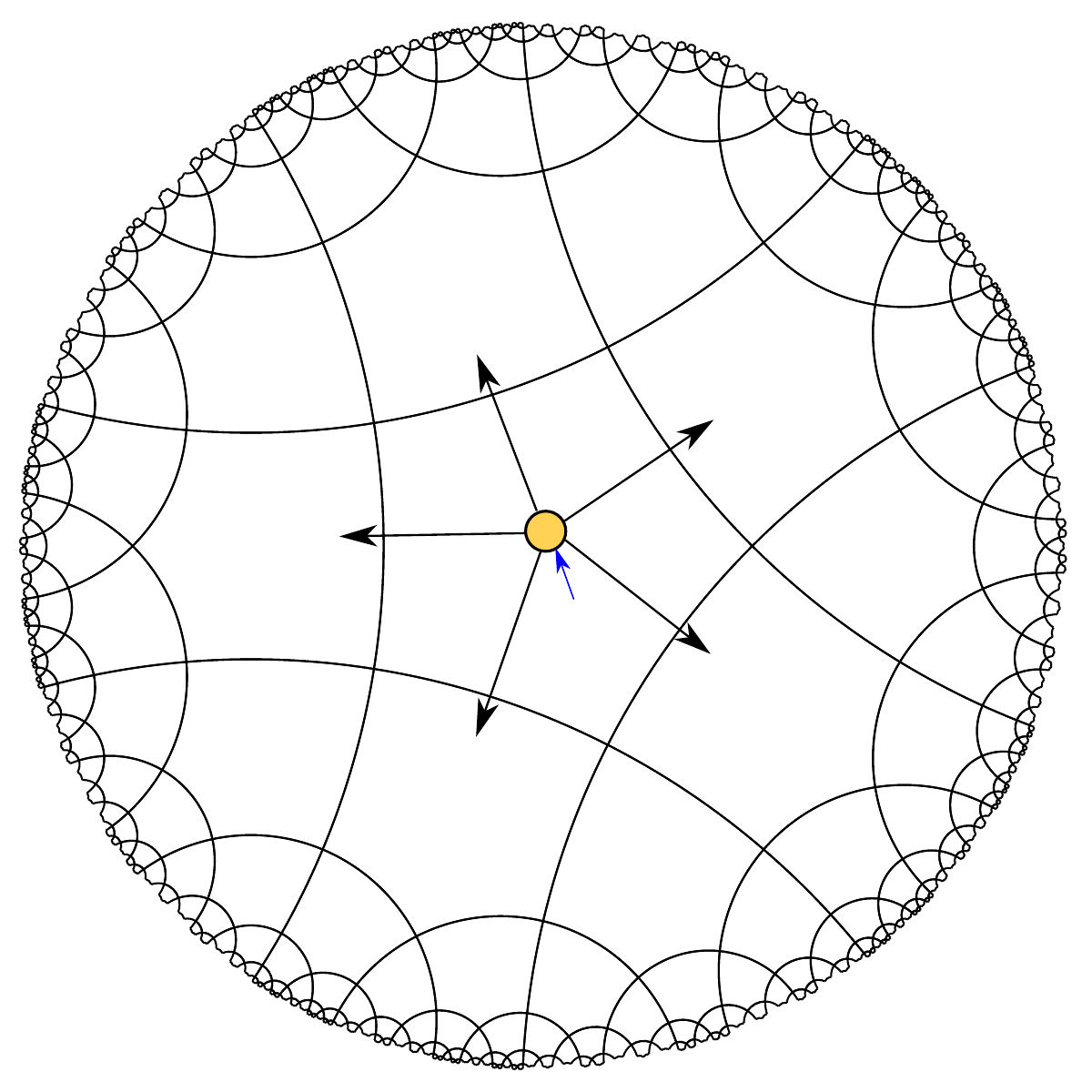}
    \caption{The $\{5, 4\}$ tiling of the hyperbolic disk. We place a six-legged perfect tensor at the center of each pentagon with one leg (show in blue) poking out of the plane of the paper.}
    \label{fig:hyperbole}
\end{figure}

\begin{figure}
\centering
    \includegraphics[width = 0.45\textwidth]{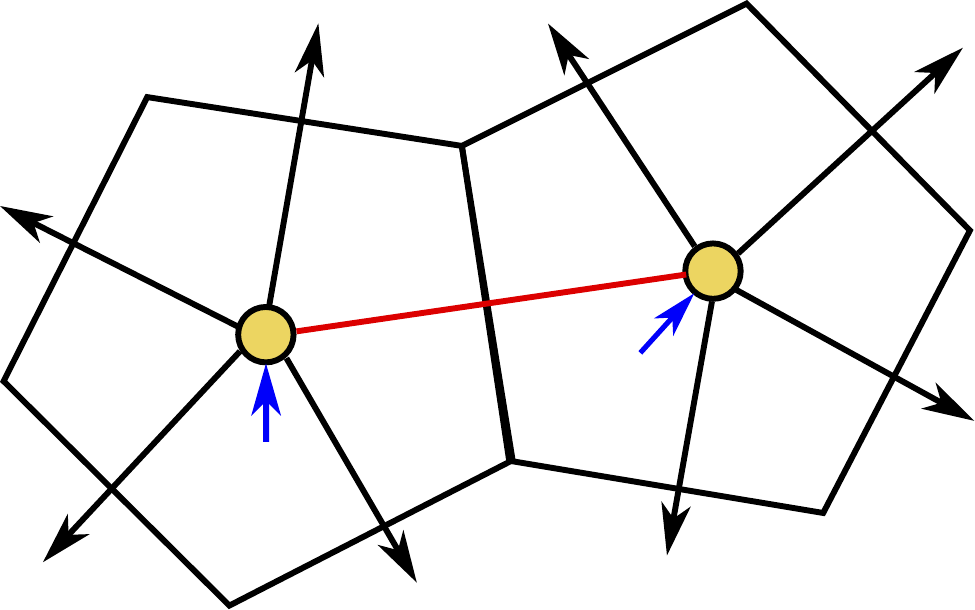}
    \caption{Contracting two legs cutting across a shared edge produces a map from $2 \to 8$ legs.}
    \label{fig:pentagon-contract}
\end{figure}

There is another version of the HaPPY code that defines a map from a set of bulk legs to the boundary legs. This is achieved by considering a tiling of the hyperbolic disk with pentagons instead of hexagons as shown in Fig. \ref{fig:hyperbole}. We again place the AME(6, 2) state at the center of each pentagon, but with a bulk leg sticking out of the paper (shown in blue). When two pentagons share an edge, the corresponding legs are contracted, and we get a map from two legs to eight legs as shown in Fig. \ref{fig:pentagon-contract}. In the next subsection, we also show that this version is a stabilizer code. The core idea is that contracting legs of a tensor maps stabilizer states to stabilizer states. 

\subsection{First version of the HaPPY code} 
It is easy to see that contracting two legs is the same as projecting onto the maximally entangled Bell state 
\begin{equation}
    \ket{\phi^+} = \frac{1}{\sqrt{N}}\sum_{i=1}^{N}\ket{ii}
\end{equation}
and tracing it out. Going back to the $T S$ example from before, consider the states 
\begin{equation}
    \ket{T} = \sum_{i_1, i_2, i_3} T_{i_1 i_2 i_3} \ket{i_1 i_2 i_3}, \quad \text{and} \quad  \ket{S} = \sum_{j_1, j_2, j_3, j_4} S_{j_1j_2j_3j_4} \ket{j_1 j_2 j_3 j_4},
\end{equation}
and consider the action of $\ket{\phi^+}\bra{\phi^+}$ on $\ket{T} \otimes \ket{S}$. The Bell pair acts on $\ket{i_1 j_3}$ giving the Kronecker delta $\delta^{i_1 j_3}$, and we are left with $\ket{\phi^+} \otimes \ket{TS}$ (up to normalization), where 
\begin{equation}
    \ket{TS} = \sum_{i_2, i_3, j_1, j_2, j_4} (TS)_{i_2 i_3 j_1 j_2 j_4} \ket{i_2 i_3 j_1 j_2 j_4}. 
\end{equation}
Tracing out the $\ket{\phi^+}$ gives us $\ket{TS}$. 

Coming back to the HaPPY code, we first observe that the two-qubit Bell state is a stabilizer state. Next, observe that a tensor product of stabilizer states is also a stabilizer state. Now comes a fact about stabilizer states that is easy to prove. Suppose $\ket{\psi}$ is a stabilizer state, and we project two of its qubits, say the $i^{\text{th}}$ and $j^{\text{th}}$ ones, onto the Bell state by acting with $\ket{\phi^+}_{ij}\bra{\phi^+}_{ij}$. In terms of an equation,
\begin{equation}\label{eq:chi}
    \left(\ket{\phi^+}_{ij} \bra{\phi^+}_{ij} \otimes \mathbf{1}\right) \ket{\psi} = \ket{\phi^+}_{ij} \otimes \ket{\chi},
\end{equation}
where $\mathbf{1}$ denotes the identity operation on the remaining qubits of $\ket{\psi}$. Then the resulting state $\ket{\chi}$ is also a stabilizer state. There's a simple way of writing down its generators using the generators of $\ket{\psi}$. Therefore, the first version of the HaPPY code prepares a stabilizer state on the boundary.

To write down the generators of $\ket{\chi}$, we proceed as follows. The projector onto the Bell state can be written as a product of the following two projectors:
\begin{equation}
\ket{\phi^+}_{ij}\bra{\phi^+}_{ij} = \frac{1}{2}(I_iI_j + X_iX_j)  \times \frac{1}{2}(I_iI_j + Z_iZ_j).
\end{equation}
Acting with $\ket{\phi^+}\bra{\phi^+}$ on a stabilizer state $\ket{\psi}$ is a two-step process. Say $\ket{\psi}$ has generators $\{g_1, \dots, g_n\}$. The first step is to find a generator, say $g_1$, which anti-commutes with $Z_i Z_j$. If there is no such element, then $Z_i Z_j$ is part of the stabilizer group of $\ket{\psi}$ and we can take $g_1 = Z_i Z_j$ to be one of the generators\footnote{Or $-Z_i Z_j$ is part of the stabilizer group of $\ket{\psi}$, in which case the projector gives zero. We will assume this does not happen. A similar condition is assumed in the second step as well.}. If there are additional generators $g_l$ that anti-commute with $Z_i Z_j$, then we replace them with $\bar{g}_l = g_1 g_l$ to ensure that only $g_1$ anti-commutes with $Z_i Z_j$. The first step ends by replacing $g_1$ with $Z_i Z_j$.

The second step is to find another generator, call it $g_2$, that anti-commutes with $X_i X_j$. If there is no such generator, then $X_i X_j$ is part of the stabilizer group of $\ket{\psi}$, and we set $g_2 = X_i X_j$. If there are additional generators $g_l$ that anti-commute with $X_iX_j$, then they are replaced by $\bar{g}_l = g_2 g_l$ as before to ensure that only $g_2$ anti-commutes with $X_iX_j$. The second step ends by replacing $g_2$ with $X_i X_j$.

At the end of these two steps, we obtain a set of $n$ generators two of which are $X_i X_j$ and $Z_i Z_j$. The generators of $\ket{\chi}$ are the remaining $n-2$ generators with their $i^{\text{th}}$ and $j^{\text{th}}$ qubits discarded. This is how we produce the generators of $\ket{\chi}$. This shows that $\ket{\chi}$ is a stabilizer state by writing down its $n-2$ generators. This argument shows that index contraction maps stabilizer states to stabilizer states.

The first version of the HaPPY code, therefore, always produces a stabilizer state on the boundary. The result from the previous section shows that such states are not holographic since they don't reproduce holographic features of tripartite entanglement. 

This concludes our discussion of the first version of the HaPPY code. Before we discuss the second version, we need to review some facts about stabilizer codes. In particular, we discuss how to think of an $[n, k]$ stabilizer code as an encoding of $k$ \emph{logical} qubits to $n$ \emph{physical} qubits. We also show that a stabilizer state on $k$ qubits is mapped to a stabilizer state on $n$ qubits under the stabilizer code. This is textbook material \cite{nielsen2010quantum}, but we nevertheless review it for the sake of completeness.

\subsection{Stabilizer codes}

So far, we looked at an $n$-qubit state that is stabilized by $n$ independent and commuting generators. We can relax this condition to define a stabilizer code. An $[n, k]$ stabilizer code is a $k$-dimensional subspace that is stabilized by $(n-k)$ independent and commuting generators. An $[n, 0]$ stabilizer code is a stabilizer state, for example.

Alternatively, we can think of a stabilizer code as a map from $k$ logical qubits to $n$ physical qubits. The image of this map is a $k$-dimensional subspace stabilized by a set of $(n-k)$ independent and commuting generators. The logical qubits are constructed by finding an orthonormal set of $2^k$ states in the code subspace to serve as a basis. The way to do this is by finding $k$ generators $\bar{Z}_1, \dots, \bar{Z}_k$ from the Pauli group $\mathcal{P}_n$ such that the $n-k$ generators of the stabilizer code, call them $\{g_1, \dots, g_{n-k}\}$, along with these $k$ generators form an independent commuting set. The logical basis state $\ket{x_1, \dots, x_k}$ is then defined to be the stabilizer state stabilized by $\{g_1, \dots, g_{n-k}, (-1)^{x_1} \bar{Z}_1, \dots, (-1)^{x_k} \bar{Z}_k\}$. Therefore, the $\bar{Z}_i$'s play the role of the logical $Z_i$ operator that maps $\ket{x_i} \to (-1)^{x_i}\ket{x_i}$. The logical $X_i$ operators $\bar{X}_i$ are defined as $\bar{X}_i \bar{Z}_j \bar{X}_i^{\dagger} = (-1)^{\delta_{ij}} \bar{Z}_j$, along with $\bar{X}_i g \bar{X}_i^{\dagger} = g$ for the generators of the code. With this construction, an $[n ,k]$ stabilizer code has the interpretation of a map from $k$ logical qubits to $n$ physical qubits defined by $Z_i \to \bar{Z}_i$ and $X_i \to \bar{X}_i$.

As a simple example, consider the three-qubit bit flip code. This is a $[3, 1]$ stabilizer code with two generators $ZZI$ and $IZZ$. To this we add $\bar{Z}_1 = ZZZ$ so that $\{ZZI, IZZ, ZZZ\}$ forms an independent commuting set. The logical qubits are $\ket{0}_L$, stabilized by the generators $\{ZZI, IZZ, ZZZ\}$, and $\ket{1}_L$, stabilized by $\{ZZI, IZZ, -ZZZ\}$. We interpret the $[3, 1]$ code as encoding one qubit into three qubits via:
\begin{equation}
 \ket{0} \to \ket{0}_L = \ket{000}, \quad \ket{1} \to \ket{1}_L = \ket{111}.
\end{equation}
The logical $X$ operator, in this case, is $\bar{X} = XXX$ whose action is $\ket{0}_L \to \ket{1}_L$ and $\ket{1}_L \to \ket{0}_L$.

Now, if $\ket{\phi}$ is a $k$-qubit stabilizer state with generators $\{g_1, \dots, g_k\}$, then an $[n, k]$ stabilizer code with generators $\{s_{k+1}, \dots, s_n\}$ maps $\ket{\phi}$ to the $n$-qubit stabilizer state $\ket{\bar{\phi}}$ with generators $\{\bar{g}_1, \dots, \bar{g}_k, s_{k+1}, \dots, s_n\}$, where $\bar{g}_i$ is defined as follows. Suppose a generator $g$ of the $k$-qubit state $\ket{\phi}$ has the form
\begin{equation}
 g = i^{r} P_1 \otimes \dots \otimes P_k
\end{equation}
where $r = 0,1,2,3$ and each $P$ is a Pauli matrix. Then,
\begin{equation}
 \bar{g} = i^r \bar{P}_1 \cdot \dots \cdot \bar{P}_k
\end{equation}
where $\bar{P}_i$ is the image of $P_i$ under the stabilizer code. From our construction, it is easy to see that the $\bar{g}$'s commute with each other and also with the $s$'s. Therefore, stabilizer codes map stabilizer states to stabilizer states. 

\subsection{HaPPY code as a stabilizer code}

In the previous subsection, we showed that an $[n, k]$ stabilizer code is a map from $k$ qubits to $n$ qubits. We also showed that any $k$ qubit stabilizer state is mapped to an $n$ qubit stabilizer state under the action of such a code.

Coming back to the second version of the HaPPY code, we think of the AME(6, 2) state as a $[5, 1]$ stabilizer code mapping one logical qubit to five physical qubits. When two pentagons share an edge, we contract with a Bell pair, and this index contraction maps stabilizer states to stabilizer states. More generally, it maps stabilizer codes to stabilizer codes. For example, the tensor network in Fig. \ref{fig:pentagon-contract} is a $[2, 8]$ stabilizer code whose generators are written down following the same logic used to write down the generators of $\ket{\chi}$ in Eq. \ref{eq:chi}. Therefore, the second version of the HaPPY code is a stabilizer code. 

As a consequence, a stabilizer state on the bulk legs is mapped to a stabilizer state on the boundary legs. Applying the result from section \ref{sec3} to this state, we violate the GHZ-forbidding inequality. There are two ways we may satisfy the GHZ-violating inequality. Either we build a tensor network using non-stabilizer AME states, or we use non-stabilizer bulk states in the second version of the HaPPY code. In either case, the boundary state is not stabilizer, and it may display features of holographic tripartite entanglement. 

\section{Discussion}
\label{sec5}

The main result of this paper, established in section \ref{sec3}, is 
\begin{equation}
 \frac{1}{2}S^{(R)}(A: B) + S(C) \leq S^{(3)}(A: B: C)
\end{equation}
for stabilizer states, in stark contrast with
\begin{equation}
 \frac{1}{2}S^{(R)}(A:B) + S(C) \geq S^{(3)}(A:B:C)
\end{equation}
for holographic states. 

Armed with the GHZ-forbidding inequality, we can shoot down stabilizer states as toy models of holographic states. In particular, we looked at two versions of the HaPPY code based on the AME(6, 2) perfect tensor. In the first version, all bulk legs were contracted, yielding a stabilizer state on the boundary. In the second version, the perfect tensor was treated as a $[5, 1]$ code giving a bulk-to-boundary stabilizer code. The tensor network prepares a stabilizer state on the boundary whenever the bulk input is a stabilizer state. Our result shows that in either case, the HaPPY code cannot capture holographic features of tripartite entanglement. This opens up a few interesting points for discussion.

\subsection*{HaPPY code without stabilizers} 

\begin{figure}
 \centering
 \includegraphics[width = 0.6\textwidth]{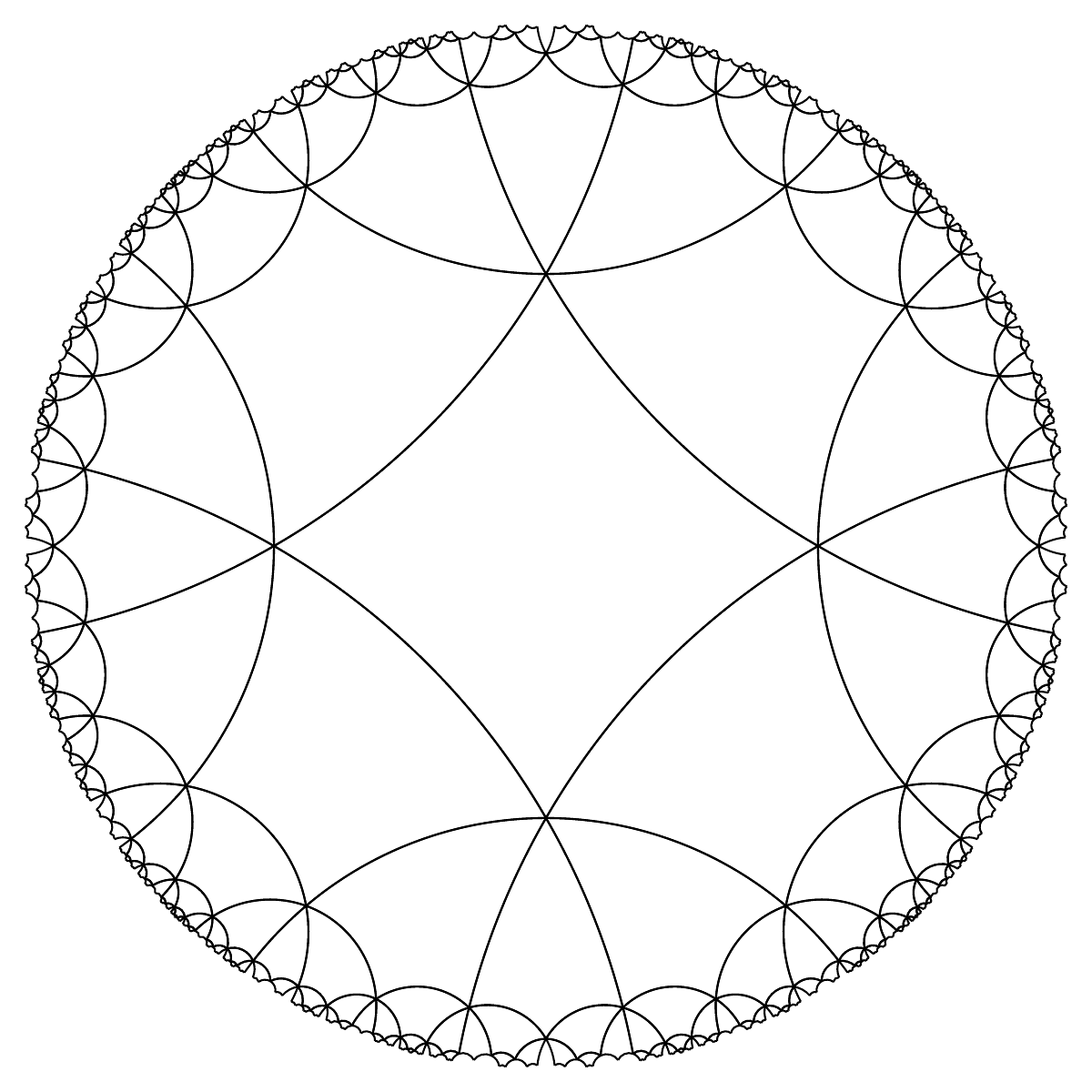}
 \caption{A hyperbolic tiling with quadrilaterals. We can place the AME(4, 6) state at the center of each quadrilateral and prepare a non-stabilizer state on the boundary. This state may reproduce holographic features of tripartite entanglement.}
 \label{fig:quadrilateral}
\end{figure}

As previously noted, there are non-stabilizer AME states. We can use these states as the building blocks of the HaPPY code to prepare non-stabilizer states on the boundary. An example of such a state is the AME(4, 6) state constructed in \cite{Rather:2021vff}. This is a four-legged perfect tensor where each leg has local dimension $6$. Consider tiling the hyperbolic disk using squares as shown in Fig. \ref{fig:quadrilateral} and place the AME(4, 6) state at the center of each square. This construction prepares a non-stabilizer boundary state that may satisfy the GHZ-forbidding inequality, and capture holographic features of tripartite entanglement.

An immediate problem in trying to check the inequality for a non-stabilizer state is the multi-entropy, which is defined as the $n \to 1$ limit of the R\'enyi multi-entropy in Eq. (\ref{eq:renyi-multi}). For general states, making sense of this limit is tricky. For example, the W state from Eq. (\ref{eq:W}) is a simple three-qubit state whose R\'enyi multi-entropy $\mathcal{Z}_n^{(3)}$ was worked out in \cite{Gadde:2023zni} for arbitrary integer values of $n$. Taking the $n \to 1$ limit, however, is highly nontrivial even for such a simple state. Unlike the entanglement entropy, the assumptions of Carlson's theorem \cite{hardy1920} don't necessarily hold for the R\'enyi multi-entropy, and it isn't clear how to make sense of the analytic continuation required to take the limit. Perhaps the limit does not even exist for general states. 

We did not face this issue for tripartite stabilizer states because the multi-entropy for the GHZ state is obtained by a straightforward $n \to 1$ limit \cite{Gadde:2022cqi}, and the multi-entropy for a Bell state is the entanglement entropy. By using the GHZ extraction theorem for, we wrote down a general formula for the tripartite multi-entropy. Understanding the $n \to 1$ limit of the R\'enyi multi-entropy is an open problem. What is the class of states for which the limit makes sense?

\subsection*{Qudits} 
Our discussion of the stabilizer formalism used qubits. The theorem we used about tripartite stabilizer states continues to hold for qudit stabilizer states \cite{Looi_2011} as long as the local dimension, $d$, is either prime or square-free. All we need to do is to replace $\log_2$ with $\log_d$ in our expressions for the reflected entropy and the multi-entropy in Eq. (\ref{eq:ref-stab}). The result from section \ref{sec3} continues to hold.  

\subsection*{Magic and complexity} 
We showed that stabilizer states cannot be holographic because they violate inequalities we expect holographic states to satisfy. In other words, holographic states are \emph{magical} \cite{Bravyi:2004isx}. There are various measures, like Gross's Wigner function \cite{Gross_2006, Gross_2006_1}, that quantify the amount of magic, or nonstabilizerness, of a quantum state. Since the GHZ-forbidding inequality rules out stabilizer states, can it be used to quantify magic? In other words, can the difference
\begin{equation}\label{eq:magic}
 \mathcal{M} = \frac{1}{2}S^{(R)}(A:B) + S(C) - S^{(3)}(A:B:C)
\end{equation}
be interpreted as a measure of magic? If so, we get a measure of magic that is basis-independent. For example, Gross's Wigner function depends on a choice of basis; it is not invariant under a local-unitary transformation. The above difference, on the other hand, is local-unitary invariant which is a desirable feature from an information-theoretic point of view. Similar ideas, stemming from a different motivation, are explored in \cite{Cao:2023mzo, Cao:2024nrx}. 

In AdS/CFT, the notion of complexity has appeared in various guises \cite{Susskind:2014rva, Brown:2015lvg, Baiguera:2025dkc}. The amount of nonstabilizerness can also be thought of as a measure of complexity. This is because stabilizer states are cheap to simulate on a classical computer. What's expensive to simulate is a non-stabilizer (magic) state. One interpretation of our result is that holographic states are complex. Our result somehow seems to be pointing to a deep relationship between multipartite entanglement and complexity/magic for holographic states. It may be that the quantity $\mathcal{M}$ defined above is, in fact, equal to some known measure of magic or complexity for holographic states. 

\subsection*{Random tensor networks} 
Another class of tensor network models that reproduce the Ryu-Takayanagi formula are random tensor networks \cite{Hayden:2016cfa}. In addition to the RT formula, they also capture the sub-leading bulk entropy contribution \cite{Faulkner:2013ana}. Furthermore, random tensor networks also reproduce the entanglement wedge cross-section for the reflected entropy \cite{Akers:2021pvd}. We suspect the multi-entropy in random tensor networks is also consistent with expectations from holography, i.e., it reproduces the minimal Mercedes-Benz. If true, then hyperbolic random tensor networks do not violate the GHZ-forbidding inequality. Another interpretation of our result is that random tensor networks are better than stabilizer tensor networks in capturing holographic features of tripartite entanglement. Multi-entropy remains to be explored in random tensor networks.

\subsection*{Four (and more) parties}

\begin{figure}
    \centering
    \includegraphics[width=0.4\linewidth]{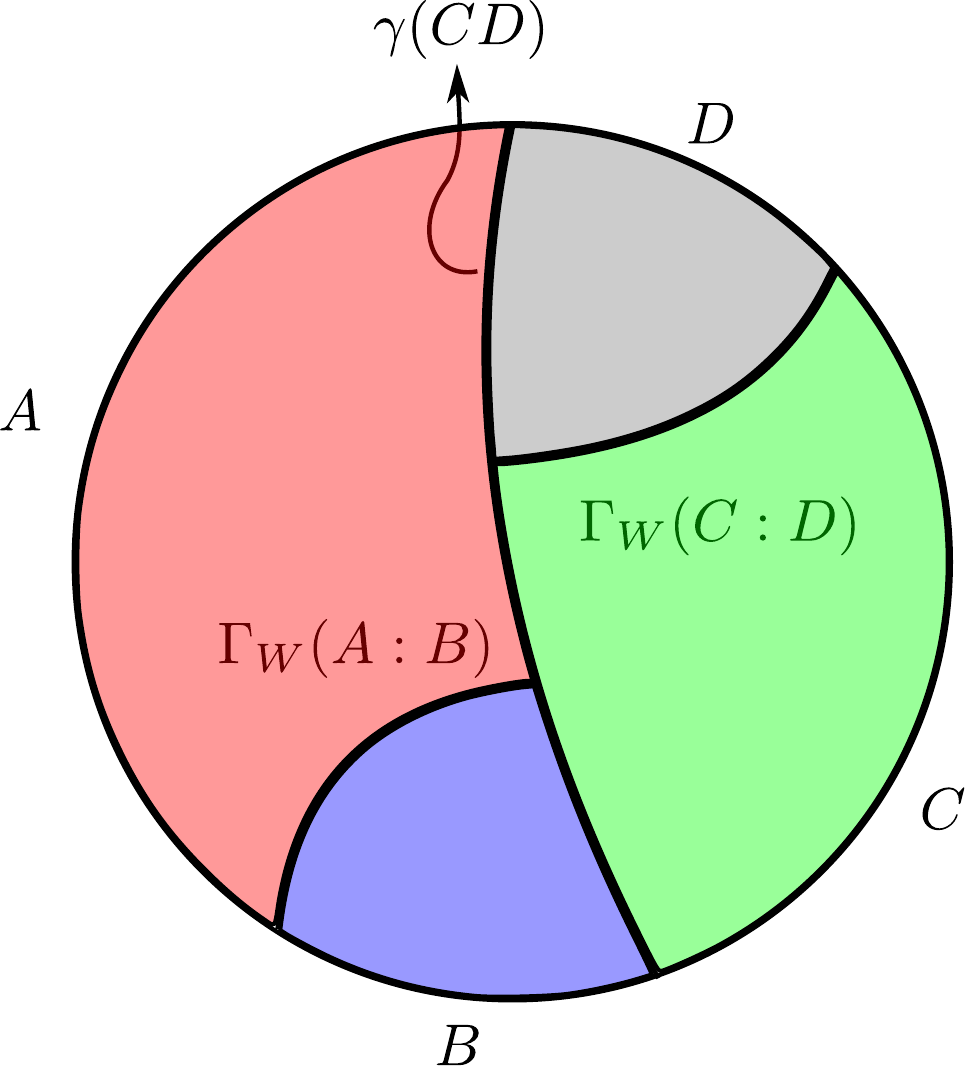}
    \caption{A bulk Cauchy slice $\Sigma$ whose boundary circle is partitioned into four parties $A$, $B$, $C$, and $D$. $\gamma(CD)$ is the Ryu-Takayanagi surface for the boundary region $C \cup D$, and $\Gamma_W(A:B)$ and $\Gamma_W(C:D)$ are entanglement wedge cross-section surfaces.}
    \label{fig:four-ew-1}
\end{figure}

Our discussion so far involved tripartite entanglement. But it isn't difficult to come up with multipartite generalizations of the GHZ-forbidding inequality. For example, consider the boundary circle again and partition it into four regions $A$, $B$, $C$, and $D$ as shown in Fig. \ref{fig:four-ew-1}. By looking at this figure, we can write down the following inequality
\begin{equation}\label{eq:4-ineq}
    \frac{1}{2}S^{(R)}(A:B) + \frac{1}{2}S^{(R)}(C:D) + S(CD) \geq S^{(4)}(A:B:C:D),
\end{equation}
where $S^{(4)}(A\!:\!B\!:\!C\!:\!D)$ is the four-partite multi-entropy. This inequality was also discussed in \cite{Balasubramanian:2025hxg} as a four-party analog of the GHZ-forbidding inequality. 

Unfortunately, we cannot check whether this inequality holds for stabilizer states or not. This is because there is no known four-party analog of the GHZ-extraction theorem. There are reasons to believe that such a classification of the four-partite entanglement structure of stabilizer states is not possible \cite{Bravyi_2006}. Without such a theorem, we have no way of calculating the four-partite multi-entropy except with a computer. There is also the pesky issue of the $n \to 1$ limit which may not exist.

Nevertheless, Eq. (\ref{eq:4-ineq}) is another holographic inequality that, in principle, probes the multipartite entanglement structure of holographic states at a deeper level than the inequalities coming from the holographic entropy cone. Plus, such inequalities are relatively easy to write down from the bulk description. There may even be a systematic way of generating such inequalities. 

In this paper, we used the tripartite GHZ-forbidding inequality to rule out stabilizer states as candidates for holographic states. We believe such multipartite holographic inequalities will continue to play a significant role in constraining the multipartite entanglement structure of holographic states. We leave these ideas for future exploration. 

\section*{Acknowledgments}
I thank my PhD advisor, Abhijit Gadde, for his support and guidance. Thanks are also due to Abhirup Bhattacharya, Ritwik Chakraborty, Shaivi Chavan, Omkar Nippanikar, Onkar Parrikar, and Rushikesh Suroshe for useful discussions and suggestions. I would also like to thank the organizers of the 31st W.E. Heraeus Summer School "Saalburg" where part of this work was completed. This work is supported by the Department of Atomic Energy, Government of India,
under Project Identification No. RTI 4002, and the Infosys Endowment for the study of the Quantum Structure of Spacetime.

\bibliography{refs}

\end{document}